\documentclass[showpacs,preprintnumbers,amsmath,amssymb,superscriptaddress]{revtex4}
\usepackage{mathrsfs}

\usepackage{amsmath,amssymb,graphicx,bm}

\begin{document}


\title{\Large \bf A Two-field Dilaton Model of Dark Energy}

\author{Nan Liang}
\email{liang-n03@mails.tsinghua.edu.cn}
 \affiliation{Department of Physics and Tsinghua Center for Astrophysics,
Tsinghua University, Beijing 100084, China}
\author{ChangJun Gao}
\email{gaocj@bao.ac.cn} \affiliation{National Astronomical
Observatories, Chinese Academy of Sciences,Beijing 100012, P. R.
China}
\author{ShuangNan Zhang}
 \email{zhangsn@mail.tsinghua.edu.cn}
 \affiliation{Department of Physics and Tsinghua Center for Astrophysics,
Tsinghua University, Beijing 100084, China} \affiliation{Key
Laboratory of Particle Astrophysics, Institute of High Energy
Physics, Chinese Academy of Sciences, P.O. Box 918-3, Beijing
100049, China}


\date{\today}

\begin{abstract}
We investigate the cosmological evolution of a  two-field model of
dark energy where one is a dilaton field with canonical kinetic
 energy and the other is a phantom field with a negative kinetic energy term.
 A phase-plane analysis shows that the phantom-dominated scaling solution is the stable late-time
attractor of this type of models. We find that during the evolution
of the universe, the equation of state $w$ changes from $w > -1$ to
$w <-1$, which is consistent with the recent observations.
\end{abstract}


\maketitle

\section{Introduction}
In recent years, observations of Type Ia supernovae (SNe Ia) [1, 2],
cosmic microwave background (CMB) fluctuations [3, 4], and
large-scale structures (LSS) [5, 6] indicate that the Universe is
accelerating, therefore some form of dark energy  whose fractional
energy density is about $\Omega_{DE}=0.70$ must exist in the
Universe to drive this acceleration. Dark energy has been one of the
most active fields in modern cosmology since the discovery of
accelerated expansion of our universe. Investigation on the nature
of dark energy becomes one of the most important tasks for modern
physics and modern astrophysics. Up to now, many candidates of dark
energy have been proposed to fit various observations which include
the simplest one, the Einstein's cosmological constant [7], or a
dynamical scalar field, such as quintessence [8], phantom [9],
k-essence [10], tachyon [11] and so on. The present data seem to
slightly favor an evolving dark energy with the equation-of-state
parameter~(EoS) $w<-1$ around present epoch and  $w>-1$ in the near
past. Obviously, $w$ cannot cross  $-1$ for quintessence or phantom
alone. Some efforts have been made to build dark energy model whose
EoS can cross the phantom divide. In a universe filled with
quintessence and phantom fields this case can be realized easily.
This implement of dark energy, called as quintom, has been first
proposed in Ref. [12], where the quintom model with an exponential
potential and the existence, stability of cosmological scaling
solutions in the context of spatially homogeneous cosmological
models have been investigated. Phase-plane analysis of the spatially
flat FRW models shows that the phantom-dominated scaling solution is
the unique late-time attractor and there exists a transition from
$w>-1$ to $w<-1$ [13]. Wei and Cai [14] suggested a hessence model,
in which a non-canonical complex scalar field plays the role of dark
energy. The cosmological evolution of the hessence dark energy is
also investigated; it is found that the big rip never appears in the
hessence model even in the most general case while beyond particular
potentials and interaction forms.

The action of dilaton field in the presence of Einstein's
cosmological constant has been derived in Ref. [15]. The potential
is the counterpart of the Einstein's cosmological constant in the
dilaton gravity theory. Since it can be reduced to the Einstein
cosmological constant when the dilaton field is set to zero, the
dilaton potential is called the cosmological constant term in the
dilaton gravity theory. Compared to the ordinary scalar field, the
action for phantom scalar field has only a sign difference before
the kinetic term. Later, the explicit expression of the phantom
potential have been given in Ref. [16]. A model of the Universe
dominated by the dilaton field with a Liouville type potential has
been presented in Ref. [17].

In this Letter, we investigate the cosmological evolution of a
two-field model of dark energy where one is a dilaton field with
canonical kinetic  energy and the other is a phantom field with a
negative kinetic energy  term with  Liouville type potentials.

\section{Equations of motion for the two-field Dilaton model}

Let us start from a 4-dimensional theory in which gravity is coupled
to dilaton and Maxwell field with an action:
\begin{equation}
S_1=\int d^4x\sqrt{-g}\left(R-2\partial _\mu \phi \partial ^\mu \phi
  -V_1(\phi)+e^{-2\alpha\phi} F^2\right),
\end{equation}
where \textsl{R} is the scalar curvature, $F^2 =
F_{\mu\nu}F^{\mu\nu}$ is the usual Maxwell contribution, $\alpha$ is
an arbitrary constant governing the strength of the coupling between
the dilaton and the Maxwell field, $V_1(\phi)$ is a potential of
dilaton $\phi$ which is given by Ref. [15]
\begin{eqnarray}
V_1(\phi)=\frac{2\lambda}{3(1+\alpha^2)}[\alpha^2(3\alpha^2-1)e^{-\frac{2}{\alpha}\phi}+
(3-\alpha^2)e^{2\alpha\phi}+8\alpha^2e^{(\alpha-\frac{1}{\alpha})\phi}],
\end{eqnarray}
here $\lambda$ is the cosmological constant. One can verify that the
potential reduces to the Einstein cosmological constant when
$\alpha=0$ or $\phi= 0$. Compared to the action of the ordinary
scalar fields, the phantom field has one negative kinetic term. In
order to obtain a real action of the Einstein-Maxwell field in the
presence of the phantom, we can make substitutions in the action as
follows Ref. [16]
\begin{eqnarray}
\phi\to i\psi,\alpha\to i\beta,
\end{eqnarray}
where $i$ is the imaginary unit. Thus we get the action
\begin{equation}
S_2=\int d^4x\sqrt{-g}\left(R+2\partial _\mu \psi \partial ^\mu \psi
  -V_2(\psi)+e^{-2\beta\psi} F^2\right),
\end{equation}
and the potential for the phantom field
\begin{eqnarray}
V_2(\psi)=\frac{2\lambda}{3(1-\beta^2)}[\beta^2(3\beta^2+1)e^{-\frac{2}{\beta}\psi}+
(\beta^2+3)e^{2\beta\psi}-8\beta^2e^{(\beta-\frac{1}{\beta})\psi}].
\end{eqnarray}
One can also verify that, when $\beta=0$ or $\psi= 0$ the action
reduces to the Einstein-Maxwell action and when $F^2=0$ the action
reduces to the Einstein-phantom action.

We consider the action in a simple model which contains a  normal
scalar field $\phi$ and a negative-kinetic scalar field $\psi$,
assuming that there is no direct coupling between the phantom field
and the normal scalar field with such  potentials,
\begin{equation}
S=\int d^4x\sqrt{-g}\left(R-2\partial _\mu \phi \partial ^\mu \phi
 +2\partial _\mu \psi \partial ^\mu \psi
 -V_1(\phi)-V_2(\psi)+\mathcal{L_{\textrm{m}}}\right),
\end{equation}
where $\mathcal{L_{\textrm{m}}}$ represents the Lagrangian density
of matter fields. Considering a flat Universe which is described by
the Friedmann-Robertson-Walker metric, the homogeneous fields $\phi$
and $\psi$ can be described by a fluid with an effective energy
density $\rho$ and an effective pressure $P$ given by
\begin{eqnarray}
\rho &=& \dot{\phi}^2-\dot{\psi}^2+\frac{1}{2}V_1(\phi)+\frac{1}{2}V_2(\psi), \\
P&=&\dot{\phi}^2-\dot{\psi}^2-\frac{1}{2}V_1(\phi)-\frac{1}{2}V_2(\psi).
\end{eqnarray}
The corresponding equation of state (EoS) parameter is given by
\begin{equation}
\label{ES}
w=\frac{\dot{\phi}^2-\dot{\psi}^2-\frac{1}{2}V_1(\phi)-\frac{1}{2}V_2(\psi)}
 {\dot{\phi}^2-\dot{\psi}^2+\frac{1}{2}V_1(\phi)+\frac{1}{2}V_2(\psi)}.
\end{equation}
Then the equations of motion for the fields and the Friedmann
equation can be written as
\begin{eqnarray}
\label{AS6} \ddot{\phi}&=&-3H\dot{\phi}-\frac{1}{4}\frac{dV_{\phi}(\phi)}{d\phi} , \\
\ddot{\psi}&=&-3H\dot{\psi}+\frac{1}{4}\frac{dV_{\psi}(\psi)}{d\psi} , \\
3H^2&=&\kappa^2\left(\dot{\phi}^2-\dot{\psi}^2+\frac{1}{2}V_1(\phi)+\frac{1}{2}V_2(\psi)+\rho_\gamma\right),
\label{AS7}
\end{eqnarray}
where $\rho_\gamma$ is the density of fluid with a barotropic
equation of state $P_\gamma=(\gamma-1)\rho_\gamma$ with $\gamma$ a
constant and $0 < \gamma \le 2$ ($\gamma=4/3$ for radiation and
$\gamma=1$ for dust mater). The equation (\ref{AS7}) is the
Friedmann constraint equation.

\section{Phase space analysis and the critical points}
In this section, we investigate the two-field Dilaton model via the
conventional phase space analysis.  Similar as in Ref. [18], we
define the following new dimensionless variables
\begin{eqnarray}
x_\phi \equiv \frac{\kappa \dot{\phi }}{\sqrt{3}H}&,& \quad y_\phi
\equiv \frac{\kappa \sqrt{V_1(\phi)}}{\sqrt{6}H},\quad
\lambda_\phi\equiv \frac{\sqrt{3}\frac{\partial _1(\phi)}{\partial
\phi}}{\kappa V_1(\phi)},\quad \Gamma_\phi\equiv
\frac{V_1(\phi)\frac{\partial^2 V_1(\phi)}{\partial^2 \phi}}
{[\frac{\partial V_1(\phi)}{\partial \phi}]^2},\nonumber \\
x_\psi\equiv \frac{\kappa \dot{\psi }}{\sqrt{3}H}&,& \quad y_\psi
\equiv \frac{\kappa \sqrt{V_2(\psi)}}{\sqrt{6}H}, \quad
\lambda_\psi\equiv \frac{\sqrt{3}\frac{\partial _1(\psi)}{\partial
\psi}}{\kappa V_2(\psi)},\quad \Gamma_\psi\equiv
\frac{V_1(\psi)\frac{\partial^2 V_1(\psi)}{\partial^2 \psi}}
{[\frac{\partial V_2(\psi)}{\partial \psi}]^2},\nonumber \\
z\equiv \frac{\kappa \sqrt{\rho_\gamma}}{\sqrt{6}H},\nonumber
\end{eqnarray}
the equations of motion (\ref{AS6})-(\ref{AS7}) can be rewritten as
the following system of equations:
\begin{eqnarray}
\label{AS1}
\frac{dx_\phi}{dN}&=&3x_\phi\left(x_{\phi}^2-x_{\psi}^2+\frac{\gamma}{2}
 z^2-1\right)+\frac{1}{2}\lambda_\phi y_{\phi}^2\,, \\
\frac{dy_\phi}{dN}&=&3y_\phi\left(x_{\phi}^2-x_{\sigma}^2+\frac{\gamma}{2}
 z^2\right)-\frac{1}{2}\lambda_\phi x_{\phi}y_{\phi}\,, \\
\frac{d\lambda_\phi}{dN} &=& -x_\phi\lambda_\psi \left(\Gamma_\phi -1\right), \\
\frac{dx_\psi}{dN}&=&3x_\psi\left(x_{\phi}^2-x_{\psi}^2+\frac{\gamma}{2}
 z^2-1\right)-\frac{1}{2}\lambda_\psi y_{\psi}^2\,, \\
\frac{dy_\psi}{dN}&=&3y_\psi\left(x_{\phi}^2-x_{\sigma}^2+\frac{\gamma}{2}
 z^2\right)-\frac{1}{2}\lambda_\psi x_{\psi}y_{\psi}\,, \\
\frac{d\lambda_\psi}{dN} &=& -x_\psi\lambda_\psi \left(\Gamma_\psi -1\right), \\
\frac{dz}{dN}&=&3z\left(x_{\phi}^2-x_{\sigma}^2-\frac{\gamma}{2}z^2-\frac{\gamma}{2}\right),
\label{AS2}
\end{eqnarray}
where  $N$ is the logarithm of the scale factor ($N\equiv \ln a$),
and the Fridemann constraint equation (11) becomes
\begin{equation}
\label{AS3} x_{\phi}^2+y_\phi^2-x_{\psi}^2+y_\psi^2+z^2=1.
\end{equation}
Different from the case of a single exponential potential, the
parameters $\lambda_{\phi,\psi}$ and $\Gamma$ here are variables of
$\phi$ and $\psi$. Strictly speaking, the above system is not an
autonomous system. Thus, if we want to discuss the phase plane, we
need to find the constraints on the potential, or equivalently the
conditions under which the potential may have the property we
require in order that we can get some explicit results.

Critical points correspond to fixed points where
$\frac{dx_\phi}{dN}=0$, $\frac{dy_\phi}{dN}=0$,
$\frac{d\lambda_\phi}{dN}=0$, $\frac{dx_\psi}{dN}=0$,
$\frac{dy_\psi}{dN}=0$, $\frac{d\lambda_\psi}{dN}=0$,
$\frac{dz}{dN}=0$. Observing these equations, one can find that the
physically meaningful critical points $(x_{\phi,c}, y_{\phi,c},
\lambda_{\phi,c}, x_{\psi,c}, y_{\psi,c}, \lambda_{\psi,c})$ of the
system are: (Note that we will restrict our discussion of the
existence and stability of critical points in the expanding
universes with $H>0$).

(i). $(\lambda_{\phi,c}\not=0,\ \lambda_{\psi,c}\not=0)$ could be
fixed by $\Gamma_\phi=1,\ \Gamma_\psi=1$;

(ii). $(x_{\phi,c}=0,\ y_{\phi, c}^2+y_{\psi,c}^2=1,\ \lambda_{\phi,
c}=0,\ x_{\psi,c}=0,\ \lambda_{\psi,c}=0)$;

(iii). $(x_{\phi, c}=0,\ y_{\phi, c}=0,\ \lambda_{\phi, c}=any,\
x_{\psi, c}=0,\ y_{\psi, c}=0,\ \lambda_{\psi,c}=any)$.

In the case of (i), when $\Gamma_\phi=1,\ \Gamma_\psi=1$, then
$\lambda_{\phi, c}=\frac{2\sqrt3}{\kappa},\ \lambda_{\psi,
c}=\frac{2\sqrt3}{\kappa}$ and the two-fields potentials are given
by
\begin{eqnarray}
V_1(\phi)&=&\lambda(e^{-\frac{2\phi}{\sqrt3}}+e^\frac{2\phi}{\sqrt3}),\\
V_2(\psi)&=&2\lambda e^{-2\psi}.
\end{eqnarray}
Thus we can determine the important parameters of the dilaton field
and the phantom field:
\begin{eqnarray}
\alpha&=&\frac{1}{\sqrt3},\ \textrm{or} \ \ \sqrt3; \ \textrm{and}\
\beta=1.\nonumber
\end{eqnarray}

In this case, the equations (\ref{AS1})-(\ref{AS2}) have one
two-dimensional hyperbola (Type $A$) embedded in four-dimensional
phase-space corresponding to kinetic-dominated solutions, with EoS
$w = 1$ and fractional energy density $\Omega_{\textrm{DE}}=1$;  a
fixed point (Type $B$) corresponding to a dilaton-dominated
solution, with $w = -1+{2}/{3\kappa^2}$ and
$\Omega_{\textrm{DE}}=1$; a fixed point (Type $C$) corresponding to
a fluid-dilaton-dominated solution, with $w = 0 $ and
$\Omega_{\textrm{DE}}={3}/{2\kappa^2}$; and a fixed point (Type $D$)
corresponding to a phantom-dominated solution, with $w =
-1-{2}/{3\kappa^2}$ and $\Omega_{\textrm{DE}}=1$ (listed in Table
1).

\begin{table}
\begin{tabular}{c c c c c c c c c} \hline
Type & $x_\phi$ & $y_\phi$ & $x_\psi$ & $y_\psi$ & $z$ & $w$&
$\Omega_{\textrm{DE}}$
 & Stability \\ \hline
$A$ & $x_{\phi}^2-x_{\psi}^2=1$ & 0 & $x_{\phi}^2-x_{\psi}^2=1$ & 0 & 0 & 1& 1& unstable \\
$B$ & $\frac{1}{\sqrt{3}\kappa}$
 & $\sqrt{ (1-\frac{1}{3\kappa^2})}$ & 0 & 0 & 0 & $-1+\frac{2}{3\kappa^2}$& 1& unstable  \\
$C$ & $\frac{\sqrt{3}\kappa}{2}$
 & $\sqrt{\frac{3\kappa^2\gamma(2-\gamma)}{4}}$& 0 & 0
 & $\sqrt{1-\frac{3\kappa^2\gamma}{2}}$ & 0& $\frac{3}{2\kappa^2}$& unstable \\
$D$ & 0 & 0 & $-\frac{1}{\sqrt{3}\kappa}$ &
 $\sqrt{ (1+\frac{1}{3\kappa^2})}$ & 0 & $-1-\frac{2}{3\kappa^2}$& 1& stable  \\\hline
\end{tabular}
\caption{The properties of the critical points in a spatially flat
FRW universe containing a phantom field and a normal scalar field in
the case of (i).}
\end{table}

In order to study the stability of the critical points, using the
Friedmann constraint equation (\ref{AS3}) we can reduce
Eqs.(\ref{AS1})-(\ref{AS2}) to four independent equations.
Substituting linear perturbations $x_\phi \to x_\phi+\delta x_\phi$,
$y_\phi \to y_\phi+\delta y_\phi$, $x_\psi \to x_\psi+\delta
x_\sigma$ and $y_\psi \to y_\psi+\delta y_\psi$ into the four
independent equations, we obtain the equation of perturbations to
the first-order:
\begin{eqnarray}
\label{AS4} \delta
x_{\phi}'&=&3\left(3x_{\phi}^2-x_{\psi}^2+\frac{\gamma}{2}
 z^2-1\right) \delta x_{\phi}+\lambda_\phi y_{\phi}\delta
 y_{\phi}-6x_{\phi}x_{\phi}\delta x_{\psi}, \\
\delta
y_{\phi}'&=&3y_\phi\left(2x_{\phi}-\frac{1}{6}\lambda_\phi\right)
\delta x_{\phi}+\left(x_{\phi}^2-x_{\psi}^2+\frac{\gamma}{2}
 z^2-1\right)\delta y_{\phi}-6x_{\phi}y_{\phi}\delta x_{\psi}, \\
\delta x_{\psi}'&=& 6x_{\phi}x_{\psi}\delta
x_{\phi}-3\left(x_{\phi}^2-3x_{\psi}^2+
\frac{\gamma}{2}z^2-1\right)\delta x_{\psi}-\lambda_\psi
y_{\psi}\delta y_{\psi}, \\
\delta y_{\psi}'&=&6x_{\phi}x_{\psi}\delta
y_{\phi}+3y_\phi(-2x_\phi-\frac{1}{6}\lambda_\psi)\delta
x_{\psi}-3\left(3x_{\phi}^2-x_{\psi}^2+\frac{\gamma}{2}z^2-1\right)\delta
y_{\psi}. \label{AS5}
\end{eqnarray}

The linear perturbations of system (\ref{AS4})-(\ref{AS5}) about
each fixed point gives four eigenvalues. The theory of stability
requires that the real part of all eigenvalues should be negative.
So we have:

Type $A$ (the kinetic-dominated solution):
\begin{displaymath}
m_1=3, \quad m_2=0, \quad m_3=3(2-\gamma),
 \quad m_4=3(1 \pm \frac{1}{\sqrt{3}\kappa}),
\end{displaymath}
indicating that this solution is always unstable.

Type $B$ (the dilaton-dominated solution):
\begin{displaymath}
m_1=\frac{6}{\kappa^2}, \quad m_2=m_3=\frac{6}{\kappa^2}-3\gamma,
\quad m_4=\frac{12}{\kappa^2}-3,
\end{displaymath}
indicating that this solution is also unstable.

Type $C$ (the fluid-dilaton-dominated solution):
\begin{displaymath}
m_1 = \frac{3\gamma}{2}, \quad m_2 = \frac{3\gamma}{2}-3,
\quad m_{3,4} = -\frac{3(2-\gamma)}{4}\left(1\pm \sqrt{1-\frac{8\gamma(\frac{2}{\kappa^2}-3\gamma)}{\frac{2}{\kappa^2}(2-\gamma)}}\right) , \nonumber \\
\end{displaymath}
indicating that this solution is still unstable.

Type $D$ (the phantom-dominated solution):
\begin{displaymath}
m_1=-\frac{1}{\kappa^2}, \quad m_2=m_3=-\frac{1}{\kappa^2}-3, \quad
m_4=-\frac{2}{\kappa^2}-3\gamma,
\end{displaymath}
indicating that this solution is stable.

In the case of (ii) $(x_{\phi, c}=0,\ y_{\phi, c}^2+y_{\psi,
c}^2=1,\ \lambda_{\phi, c}=0,\ x_{\psi ,c}=0,\ \lambda_{\psi,
c}=0)$, the equations (\ref{AS1})-(\ref{AS2}) have one fixed point
Type $E$ embedded in six-dimensional phase-space which corresponds
to eigenvalues $(-3, -3\gamma, 0, 0, 0, 0)$,  $w = -1$ and
$\Omega_{\textrm{DE}}=1$. This indicates that the critical point is
a de Sitter attractor.

In the case of (iii) $(x_{\phi, c}=0,\  y_{\phi ,c}=0,\
\lambda_{\phi, c}= any,\ x_{\psi, c}=0,\ y_{\psi, c}=0,\
\lambda_{\psi, c}=any)$, the equations (\ref{AS1})-(\ref{AS2}) have
one fixed point Type $F$ embedded in six-dimensional phase-space
which corresponds to eigenvalues $(-3(\gamma-2)/2, -3\gamma/2, 0,
-3(\gamma-2)/2, -3\gamma/2,0)$,  $w$ becomes meaningless and
$\Omega_{\textrm{DE}}=0$. This indicates that the critical point is
not a dynamical attractor.

\section{Numerical Studies}

Our numerical studies indicate that the EoS parameter $w$ changes
from $>-1$ to $<-1$  as shown in Figure 1. We have assumed that
there is no direct coupling between the phantom field and the normal
scalar field in this paper. Without the loss of generality, the
initial conditions $\phi(0),\ \psi(0),\ \dot{\phi}(0)$ and
$\dot{\psi}(0)$ can be fixed in order to get the EoS today
$(w=-1.02, \ a=1)$ [19], and the energy density of dark energy today
$\Omega_{DE}=0.70$. The blue line represents the EoS of the
two-field dilaton model, the black line represents the single
dilaton model and the red line represents the single phantom model.

\begin{figure}
\begin{center}
\includegraphics[scale=1]{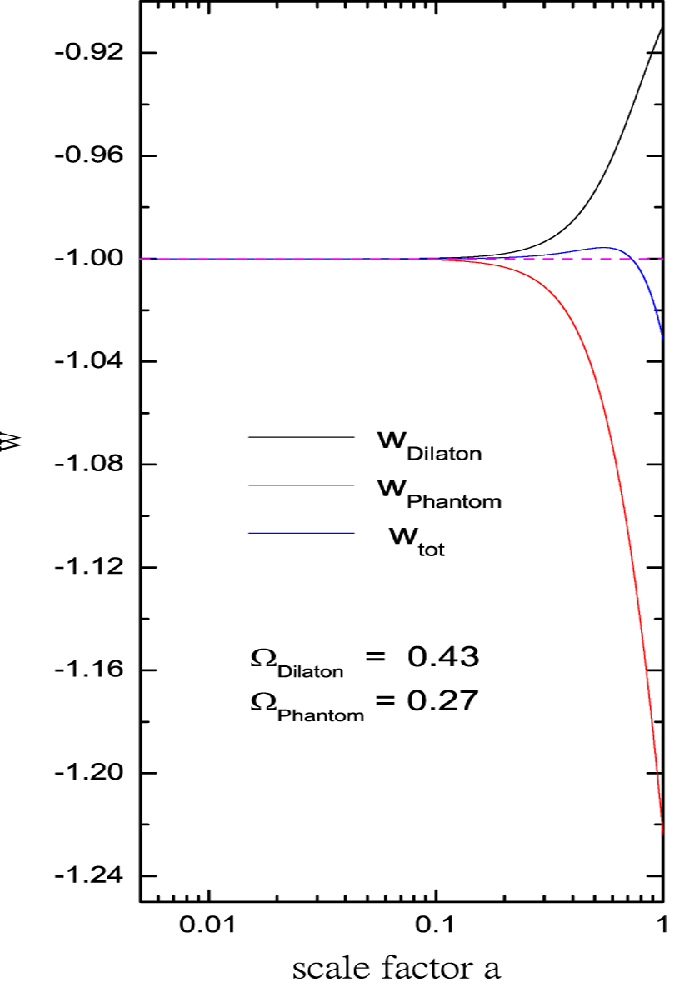}
\caption{The evolution of the equation of state $w$ of the two-field
dilaton model of dark energy. The blue line represents the EoS of
the two-field model, the black line represents the single dilaton
model and the red line represents the single phantom model.}
\end{center}
\end{figure}

\section{Conclusions and Discussions}

In summary, we have investigated the possibility of constructing a
two-field dark energy model  which has the equation of state $w$
crossing $-1$ by using the dilaton and phantom fields. We have made
a phase-space analysis of the evolution for a spatially flat FRW
universe filled with a barotropic fluid and phantom-dilaton fields.
It is shown that there exists the stable late-time attractor
solution in the model. Also, we showed that the equation of state
$w$ can cross $-1$ naturally. So the two-field dilaton field is a
viable candidate for dark energy.

It is apparent that our model is also plagued with the instability
problem at the quantum level which makes its existence doubtful. In
fact, this is a common problem for nearly all phantom models.
However, as argued by Carroll et al. [20], these models might be
phenomenologically viable if considered as effective field theories
valid only up to a certain momentum cutoff. According to their
discussions, the instability timescale of the phantom quanta can be
greater than the age of the universe provided that the cutoff is at
or below 100 MeV. In this sense, the phantom quanta are stable
against decay into gravitons and other particles. Therefore,
considering astronomical observations favoring the phantom model for
dark energy, it remains open if the phantom matter exists and acts
as dark energy.

\begin{acknowledgments}
We are grateful to Zhang Xin-Min, Wei Hao, Feng Bo for kind help and
discussions. We also thank Zhao Gong-Bo~for plotting the Fig. 1.
This project was in part supported by the Ministry of Education of
China, Directional Research Project of the Chinese Academy of
Sciences under project KJCX2-YW-T03, and by the National Natural
Science Foundation of China under grants 10521001, 10733010, and
10725313, and by 973 Program of China under grant 2009CB824800.

\end{acknowledgments}


\clearpage

\end{document}